\documentclass[12pt]{article}
\usepackage{amssymb,cite,graphics}

\title{The Influence of Strong Magnetic Field on Photon-neutrino Reactions}

\author{{M.V.~Chistyakov \thanks{E-mail address:mch@univ.uniyar.ac.ru}, N.V.~Mikheev \thanks{E-mail address:mikheev@univ.uniyar.ac.ru}}\\ [7mm] 
{\small\it Division of Theoretical Physics, Department of Physics,}\\
{\small\it Yaroslavl State University, Sovietskaya 14,}\\
{\small\it 150000 Yaroslavl, Russian Federation}}

\newcommand{\prp}[1]{#1_{\mbox{\tiny $\bot$}}}
\newcommand{\prl}[1]{#1_{\mbox{\tiny $\|$}}}
\def\th{\rm tanh}
\def\ee{\varepsilon}
\def\HH{H\!\!\left(\frac{4 m_e^2}{\prl{q}^2}\right)}

\date{}

\begin{document}


\maketitle

\thispagestyle{empty}

\begin{abstract}
{
The two-photon two-neutrino interaction induced by magnetic field 
is investigated. In particular  the processes $\gamma \gamma \to \nu \bar \nu$
and $\gamma \to \gamma  \nu \bar \nu$   
are studied in the presence of strong magnetic field. An effective 
Lagrangian and partial amplitudes of the processes are presented.
Neutrino emissivities due to the reactions $\gamma \gamma \to \nu \bar \nu$ 
and $\gamma \to \gamma  \nu \bar \nu$ 
are calculated  taking into account of the photon dispersion and large 
radiative  corrections. A comparison of the results obtained with previous 
estimations and another inducing mechanisms of the processes under 
consideration is made.  
\\
\\
{{\bf Key words:} photon-neutrino processes, electron propagator, 
magnetic field, photon dispersion}
\\
\\
PACS numbers: 12.20.Ds, 95.30.Cq, 14.60.Lm

}
\end{abstract}

\section{Introduction}

Historically the reaction $\gamma \gamma \to \nu \bar \nu$  was one of 
the first photon-neutrino processes considered in the context of 
its astrophysical application. 
In 1959 Pontecorvo suggested that $(e \nu) (e \nu)$ coupling 
could induce reactions leading to energy loss in stars~\cite{Pontecorvo:1959}. 
One of these processes, $\gamma \gamma \to \nu \bar \nu$, caused by 
this coupling  
was compared in~\cite{Chui:1960}  
with other neutrino reactions and a rough estimation of the neutrino 
energy loss rate was obtained. In both papers the authors used the 
four-fermion (V-A) Fermi model. However, in 1961 it was proved that in this case the 
process under consideration is forbidden. 
This statement is also known as the Gell-Mann theorem~\cite{Gell-Mann:1961} 
asserts that for massless neutrino and on-shell photons, in the local 
limit of weak interaction, the amplitude of the $\nu \nu \gamma 
\gamma$-interaction is equal to zero. Really, in the center-of-mass 
frame the the left neutrino and right antineutrino carry out the total 
momentum unity in the local limit of the weak interaction. However,  the 
system of two photons can't exist in the state with the total angular 
momentum equals to unity (the Landau-Yang theorem~\cite{Landau:1948, Yang:1950}).
This statement also can be formulated in another way. 
The most general amplitude of the process can be written in the gauge 
invariant form:
\begin{eqnarray}
{\cal M} = {\alpha \over \pi} \, {G_{\mbox{\normalsize{F}}} \over \sqrt{2}}\,
\left [\bar\nu_i (k_1) \,
T_{\alpha \beta \mu \nu} \,
\nu_i (- k_2) \right ]\,
f^{\alpha \beta}_1 f^{\mu \nu}_2, 
\label{Mgen}
\end{eqnarray}
where $i$ is the neutrino flavour, $i = e, \mu, \tau$, and the tensors  
$f^{\alpha \beta} = q^\alpha \varepsilon^\beta - q^\beta \varepsilon^\alpha$
are the photon field tensors in the momentum space. The Gell-Mann theorem 
means that in the case when above-listed conditions are realised 
it is impossible to  construct the nonzero tensor $T_{\alpha \beta \mu 
\nu}$.
Any deviation from the Gell-Mann 
theorem conditions leads to the nonzero amplitude of the process. For 
example, in the case  of massive neutrino the process is allowed due to 
the change of the neutrino helicity~\cite{Crewther:1982,Dodelson:1991}. The tensor $T_{\alpha \beta \mu 
\nu}$ has the following form in this case:
\begin{eqnarray}
T_{\alpha \beta \mu \nu} = 
- {i \over {24}} \, 
\left (2 \delta_{i e} - 1 \right ) \,
\frac{m_{\nu_i}} {m^2_e} \, \gamma_5 \, 
\varepsilon_{\alpha \beta \mu \nu}.
\end{eqnarray}

In the case of non-locality of the weak interaction the neutrino momenta, 
$k_{1 \mu}, k_{2 \mu}$, 
become separated and the following structure 
arises~\cite{Levine:1967,Dicus:1972,Dicus:1993}:
\begin{eqnarray}
T_{\alpha \beta \mu \nu} = {8 i \over 3} \left (\ln{m^2_W \over m^2_e}
+ {3 \over 4} \right ) {1 \over m^2_W} 
\left [\gamma_\alpha \, g_{\beta \mu} (k_1 - k_2)_\nu + 
\gamma_\mu \, g_{\nu \alpha} (k_1 - k_2)_\beta \right ]
 (1 + \gamma_5). 
\label{eq:Tnl}
\end{eqnarray}
It is seen that in both cases the amplitude is suppressed, either by 
small neutrino mass in the numerator or by large $W$-boson mass in the 
denominator, and the contribution of this channel into the stellar 
energy-loss appear to be small.

One more exotic case of nonzero  amplitude is realised for off-shell 
photons~\cite{Cung:1975,Kuznetsov:1993a}, $q_\mu f^{\mu \nu} \neq 0$, when the photon momenta can 
be included into the tensor $T_{\alpha \beta \mu \nu}$
\begin{eqnarray}
T_{\alpha \beta \mu \nu} = 
- {i \over {24}} \, 
\left (2 \delta_{i e} - 1 \right ) \,
{1 \over m^2_e} \, \gamma^\rho (1 + \gamma_5) 
 \left (\varepsilon_{\rho \alpha \mu \nu} \, q_{1 \beta}
\; 
+ \varepsilon_{\rho \mu \alpha \beta} \, q_{2 \nu}
\right ) 
\end{eqnarray}

However, there exist one more possibility to construct 
the nonzero amplitude (\ref{Mgen}). In the presence of the external 
electromagnetic field an additional tensor of electromagnetic field 
$F_{\mu \nu}$ arises which allows one to construct the tensor 
$T_{\alpha \beta \mu \nu}$ even in the case when Gell-Mann theorem 
conditions are realised. Therefore, one could say that external 
electromagnetic filed  induces the two photons two neutrinos interaction. 

Regarding possible astrophysical application of 
the process discussed, it is interesting to consider the magnetic field case. 
It is well known that magnetic field presents practically in all 
astrophysical environments. In many of them the existence of intense 
magnetic field is assumed. For example, the typical magnetic filed of 
neutron stars is observed about $10^{12}$ G. Some modern neutron star models 
consider the generation of magnetic fields up to 
$10^{14} \div 10^{16}$ G~\cite{mag}. Moreover, the very recent 
observations of SGR and AXP pulsars indicate the existence of the 
magnetic field about $10^{15}$ G~\cite{magobsv}.   
Note that the strength of such magnetic 
fields exceeds essentially the so-called critical value 
$B_e  = e / m_e^2 \simeq 4.41 \cdot 10^{13}$ G,
which is a natural scale for the field strength~
\footnote{We use the natural units, $c = \hbar = k = 1$, hereafter $e > 0$ is 
an elementary charge.}.
It is known that strong magnetic field could enhance the processes 
suppressed in vacuum (see e.g. \cite{Mikheev nng}),
therefore it could be important to investigate the influence of strong 
external magnetic field on the process $\gamma \gamma \to \nu \bar \nu$.

\section{The process $\gamma \gamma \to \nu \bar \nu$ in external magnetic field}

Previously the process under consideration was studied in the relevantly 
weak magnetic field, $B \ll B_e$. In the paper~\cite{Shaisultanov:1998} 
an effective Lagrangian 
of the $\gamma \gamma \gamma \nu \nu$-interaction~\cite{Dicus:1997} was used 
to obtain the cross 
section and the emissivity of the process $\gamma \gamma \to \nu \bar \nu$ 
with photon and neutrino energies much less than the electron mass. 
It was shown that the cross section of the process is 
enhanced by the factor $(m_W / m_e)^4 (B / B_e)^2$ in comparison with  
its counterpart in vacuum, where $m_W$ and $m_e$ are W-boson and 
electron masses respectively. Another approach was developed 
in~\cite{Chyi:1999, Chyi:2000}, where 
an electron propagator expansion in powers of the magnetic 
field  strength was applied to study the process 
$\gamma \gamma \to \nu \bar \nu$ with energies greater than $m_e$. In the 
low-energy limit the amplitude of the process obtained 
in~\cite{Chyi:1999,Chyi:2000} 
agrees with the result of Ref.~\cite{Shaisultanov:1998}. 
In the paper~\cite{Dicus:2000} the 
results~\cite{Shaisultanov:1998} and~\cite{Chyi:1999,Chyi:2000} 
were slightly corrected. 
In particular, it was noted that the cross section of the process 
$\gamma \gamma \to \nu \bar \nu$  has to be less by factor $4 \pi$.

An investigation of the low energy two photon neutrino interaction in strong 
magnetic field was performed in ~\cite{Loskutov:1981}. The amplitude and 
emissivity of the reaction $\gamma \gamma \to \nu \bar \nu$ was obtained 
in the four-fermion model without Z-boson contribution.

The purpose of this paper is to investigate the two-photon two-neutrino 
processes in the presence of strong magnetic 
field with energies restricted only by the value of the 
magnetic field strength, $\omega \ll \sqrt{eB}$. 
These processes are considered in the framework of the
Standard Model using an  effective local Lagrangian of the 
neutrino-electron interaction
\begin{equation}
{\cal L} \, = \, \frac{G_F}{\sqrt 2}\,
\big [ \bar e \gamma_{\alpha} (g_V - g_A \gamma_5) e \big ] \,
j_{\alpha} \,,
\label{L}
\end{equation}
where $g_V = \pm 1/2 + 2 \sin^2 \theta_W, \, g_A = \pm 1/2$. Here 
the upper signs correspond to the electron neutrino ($\nu = \nu_e$) 
when both Z and W boson exchange takes part in a process.  The low signs 
correspond to $\mu$ and $\tau$ neutrinos ($\nu = \nu_\mu, \nu_\tau$) when 
Z boson exchange is only presented in the Lagrangian (\ref{L}), 
$j_{\alpha} = \bar \nu \gamma_\alpha (1 - \gamma_5) \nu$ is the left neutrino 
current.

\begin{figure}[htb]
\centerline{\includegraphics{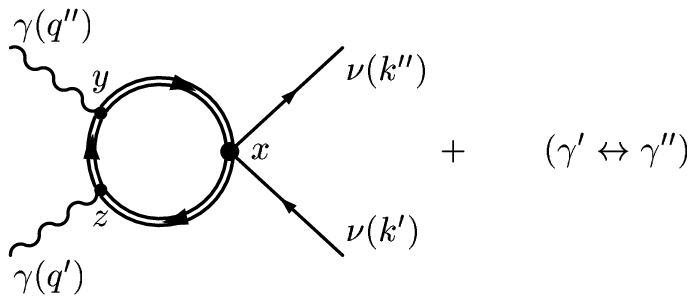}}
\caption{
The  Feynman diagram for the $\nu \nu \gamma \gamma$-interaction in 
magnetic field.}
\label{F1}
\end{figure}

In the third order of the perturbation theory the process 
$\gamma \gamma \to \nu \bar \nu$ is described by two Feynman diagrams 
depicted in Fig.~\ref{F1}, where double lines imply  that the influence of the 
external filed in the propagators of  electrons is taken into account 
exactly.

The general form for the matrix element corresponding to diagrams in 
Fig.~\ref{F1}
is the following
\begin{eqnarray}
{\cal S} &=& \frac{i 4 \pi \alpha G_{F}/\sqrt{2}}
{\sqrt{2 E' V\,2 E'' V\, 2\omega' V\, 2\omega'' V}}
\int d^4 x\, d^4 y\, d^4 z\, Sp \{(j\gamma) (g_V - g_A \gamma_5) 
S(x,y) 
\times \nonumber \\ 
&\times& \ 
(\ee''\gamma) S(y,z)  (\ee'\gamma) S(z,x)\} 
e^{-i(kx - q'z - q''y)} + (\ee', q' \leftrightarrow \ee'', q''),
\label{S_gen}
\end{eqnarray}
where $\alpha \simeq 1/137$ is the fine-structure constant; 
$q'_\alpha = (\omega', {\bf q'}), \, q''_\alpha = (\omega'', {\bf q''})$ 
are the four-momenta of the initial photons with polarisation vectors 
$\ee'_\alpha, \ee''_\alpha$ respectively; $k_\alpha$ is the 
neutrino antineutrino pair four-momentum.  
$S(x,y)$ is the electron 
propagator in the magnetic field which could be presented in the 
form
\begin{eqnarray}
S(x,y) &=& e^{i \Phi(x,y)} \hat S(x-y),
\label{S}
\\
\Phi(x,y) &=& - e \int \limits_x^y d \xi_\mu \left[ A_\mu(\xi) + 
\frac{1}{2} F_{\mu \nu} (\xi - y)_\mu \right ],
\label{Phi}
\end{eqnarray}
where $A_\mu$ and $F_{\mu \nu}$ are 4-potential and tensor of the 
uniform magnetic filed correspondingly. 
The translational invariant part $\hat S(x - y)$ has different 
representations. For our purpose it is convenient to take it in 
the following form 
\begin{equation}
\hat S(X) = \hat S_{-}(X) + \hat S_{+}(X) + \hat \prp{S}(X),
\end{equation}
where 
\begin{eqnarray}
\hat S_{\pm}(X) &=& - \frac{i}{4 \pi} \int \limits^{\infty}_{0} 
\frac{d\tau}{\th \tau}\,
\int \frac{d^2 p}{(2 \pi)^2}
[\prl{(p \gamma)} + m]\Pi_{\pm}(1 \mp \th \tau) \times
\nonumber \\
&\times& \exp\left(- \frac{eB\, \prp{X}^2}{4\, \th \tau} - 
\frac{\tau(m^2 - \prl{p}^2)}{eB} - i \prl{(pX)} \right),
\label{Spm}
\\[3mm]
\hat \prp{S}(X) &=&
- \frac{eB}{8 \pi} \int \limits^{\infty}_{0} 
\frac{d\tau}{\th^2 \tau}\,
\int \frac{d^2 p}{(2 \pi)^2}
 \prp{(X \gamma)} (1 - \th^2 \tau) \times
\nonumber \\
&\times& \exp\left(- \frac{eB\, \prp{X}^2}{4\, \th \tau} - 
\frac{\tau(m^2 - \prl{p}^2)}{eB} - i \prl{(pX)} \right).
\label{Sp}
\\
d^2 p &=& dp_0 dp_3, \quad \Pi_{\pm} = \frac{1}{2} (1 \pm i \gamma_1 
\gamma_2), \quad 
\Pi^2_\pm = \Pi_\pm, \quad [\Pi_\pm, \prl{(A \gamma)}] = 0.
\nonumber
\end{eqnarray}
 
\noindent Here $\gamma_\alpha$ are the Dirac matrices in the standard 
representation, the four-vectors with the indices $\bot$ and $\parallel$ 
belong 
to the Euclidean (1, 2) subspace and the Minkowski (0, 3) subspace correspondingly, when the 
field $\bf B$ is directed along the third axis. Then for arbitrary 4-vectors 
$A_\mu$, $B_\mu$ one has
\begin{eqnarray}
\prp{A}^\mu &=& (0, A_1, A_2, 0), \quad  \prl{A}^\mu = (A_0, 0, 0, A_3), \nonumber \\
\prp{(A B)} &=& (A \Lambda B) =  A_1 B_1 + A_2 B_2 , \quad 
\prl{(A B)} = (A \widetilde \Lambda B) = A_0 B_0 - A_3 B_3, \nonumber 
\end{eqnarray}

\noindent where the matrices 
$\Lambda_{\mu \nu} = (\varphi \varphi)_{\mu \nu}$,\,  
$\widetilde \Lambda_{\mu \nu} = 
(\tilde \varphi \tilde \varphi)_{\mu \nu}$ are constructed with 
the dimensionless tensor of the external 
magnetic field, $\varphi_{\mu \nu} =  F_{\mu \nu} /B$, 
and the dual one,
${\tilde \varphi}_{\mu \nu} = \frac{1}{2} 
\varepsilon_{\mu \nu \rho \sigma} \varphi_{\rho \sigma}$. 
Matrices $\Lambda_{\mu \nu}$ and  $\widetilde \Lambda_{\mu \nu}$
are connected by the relation 
$\widetilde \Lambda_{\mu \nu} - \Lambda_{\mu \nu} = 
g_{\mu \nu} = diag(1, -1, -1, -1)$, 
and play the role of the metric tensors in perpendicular ($\bot$)
and  parallel ($\parallel$) subspaces respectively.

In spite of the translational and gauge noninvariance of the phase 
$\Phi(x, y)$ in the propagator (\ref{S}), the total phase of three propagators 
in the loop of Fig.1 is translational and gauge invariant
\begin{eqnarray} 
\Phi(x, y) + \Phi(y, z) + \Phi(z, x) = - \frac{e}{2} (z - x)_\mu 
F_{\mu \nu} (x - y)_\nu. \nonumber
\end{eqnarray} 
This fact allows one to define the amplitude of the process in the standard 
manner
\begin{eqnarray}
{\cal S} = \frac{i (2 \pi)^4 
\delta^4 (k - q' - q'')}
{\sqrt{2 E' V\,2 E'' V\, 2\omega' V\, 2\omega'' V}}\,
{\cal M}, 
\label{Lambda_def}
\end{eqnarray}
where the amplitude ${\cal M}$ can be presented in the following 
form
\begin{eqnarray}
{\cal M} = \frac{G_{F}}{\sqrt{2} \ e} \,j_\mu \ee''_\nu \ee'_\rho	
\{g_V \Pi^V_{\mu\nu \rho}  - g_A \Pi^A_{\mu\nu \rho}\},
\label{M}
\end{eqnarray}
\begin{eqnarray}
\Pi^V_{\mu\nu \rho} &=& e^3 
\int d^4 X\, d^4 Y\, Sp \{\gamma_\mu 
 \hat S(X) \gamma_\nu 
 \ \hat S(-X-Y)  \gamma_\rho \hat S(Y)\} \times
\nonumber \\
& \times &e^{- i e\,(X F Y)/2}\; e^{i (q' X - q'' Y)} + 
(\ee', q' \leftrightarrow \ee'', q''),
\\
\Pi^A_{\mu\nu \rho} &=& e^3 
\int d^4 X\, d^4 Y\, Sp \{\gamma_\mu \gamma_5 
\hat S(X) \gamma_\nu \ \hat S(-X-Y)  \gamma_\rho \hat S(Y)\} \times
\nonumber \\
& \times &
e^{- i e\,(X F Y)/2}\; e^{i (q' X - q'' Y)} + 
 (\ee', q' \leftrightarrow \ee'', q''),
\end{eqnarray} 
with $X = z - x, \, Y = x - y$. 

In a general case substitution of the propagator (\ref{S}) into the 
amplitude (\ref{M}) leads to 
a very cumbersome expression in the form of the triple integral over the 
proper time. It is advantageous to use the asymptotic expression of the 
electron propagator for an analysis of the  amplitude in strong 
magnetic field. This asymptotic could be derived from 
Eqs.(\ref{S})-(\ref{Sp}) in the limit~$eB/\vert m^2~-~\prl{p}^2~\vert~\gg~1$. In this case the  
parts $\hat S_{\pm},\, \hat \prp{S}$ of the propagator take the form
\begin{eqnarray}
\!\!\!\!\!\!\!\!\!\!\!\!\!\!\!\!\!\!\!\!\!&&\hat S_{-}(X) \simeq 
\frac{i eB}{2 \pi} \exp(- \frac{eB \prp{X}^2}{4}) 
\int \frac{d^2 p}{(2 \pi)^2}\, \frac{\prl{(p\gamma)} + m}{\prl{p}^2 - m^2}
\Pi_{-}e^{-i \prl{(pX)}}, 
\label{S-a}
\\
\!\!\!\!\!\!\!\!\!\!\!\!\!\!\!\!\!\!\!\!\!&&\hat S_{+}(X)  \simeq 
-\frac{i}{4 \pi}\, 
[i\prl{(\gamma \,\partial/\partial X)} + m]\, 
\prl{\delta}^2(X) \, \Pi_{+} \, 
\exp(\frac{eB \prp{X}^2}{4})\, 
\Gamma (0, \frac{eB \prp{X}^2}{2}),
\label{S+a}
\\
\!\!\!\!\!\!\!\!\!\!\!\!\!\!\!\!\!\!\!\!\!&&\hat \prp{S}(X)  \simeq 
-\frac{1}{2 \pi} \, \prl{\delta}^2(X)\, \frac{\prp{(X \gamma)}}{\prp{X}^2}
\exp(- \frac{eB \prp{X}^2}{4}),
\label{S_pr_a} 
\end{eqnarray} 
where $\Gamma(a, z)$ is the incomplete gamma function
$
\Gamma(a,z) = \int \limits^\infty_z t^{a - 1} e^{-t} dt.
$

By using, (\ref{S-a})-(\ref{S_pr_a}) the amplitude ${\cal M}$ can be 
presented as a sum of the ten independent parts which can conditionally be divided into 
four groups: 
1)~$\hat S_{\pm}\hat S_{\pm}\hat S_{\pm}$;
2)~$\hat S_{\pm}\hat S_{\pm}\hat \prp{S}$; 
3)~$\hat S_{\pm}\hat \prp{S}\hat \prp{S}$; 
4)~$\hat \prp{S}\hat \prp{S}\hat \prp{S}$.  
Analysing these combinations one could expect that the 
leading on field strength part of the amplitude, namely $\sim eB$, 
arises from the combination 
$\hat S_{-}\hat S_{-}\hat S_{-}$. 
Really, substitution of $\hat S_{-}$ in amplitude (\ref{M}) gives $(eB)^3$ in 
numerator, while integration over transverse coordinates $\prp{X}$ leads to 
the factor $(eB)^2$ in denominator. Then expending the amplitude in powers 
of inverse magnetic field strength B we obtain the linear on field dependence. 
However, it is easy to see that 
two parts 
of the amplitude (\ref{M}) with photons exchange 
$(\ee', q' \leftrightarrow \ee'', q'')$ 
cancel each other exactly in this limit.
Hence the amplitude of the process $\gamma \gamma \to \nu \bar \nu$ 
doesn't depend on B in the strong magnetic field.

The analysis shows that the independent on field contribution to the 
amplitude is given by the combinations 
$\hat S_{-}\hat S_{-}\hat S_{+},\, \hat S_{-}\hat S_{-}\hat \prp{S}$ and
$\hat S_{-}\hat \prp{S} \hat \prp{S}$
with all interchanges.
One more contribution comes from the expansion of the 
$\hat S_{-}\hat S_{-}\hat S_{-}$ combination in the powers of inverse 
magnetic field strength $B$. Then substituting (\ref{S-a})-(\ref{S_pr_a}) into 
(\ref{M}) we obtain the following result for the amplitude
\begin{eqnarray}
{\cal M} \simeq \frac{G_{F}}{\sqrt{2} \ e} \,j_\mu \ee''_\nu \ee'_\rho	
\{g_V \Pi^V_{\mu\nu \rho}  - g_A \Pi^A_{\mu\nu \rho}\},
\label{M_fin}
\end{eqnarray}
\vspace*{-5mm}
\begin{eqnarray}
 \Pi^V_{\mu\nu \rho} &=&- \frac{ i e^3}{2 \pi^2} \{ (q' \varphi q'') \, 
\pi_{\mu\nu \rho} + 
 (q' I'')_\nu \,\varphi_{ \rho \mu} 
+ \frac{1}{2} ((q'' - q') I )_\mu\, 
\varphi_{ \nu \rho}  
\nonumber \\[3mm]
&+&
(q'' I')_\rho \,\varphi_{ \nu \mu} - 
 I''_{\nu \rho}\, (q' \varphi)_\mu + 
 I''_{\mu \nu}\, (q \varphi)_\rho
+ I'_{\mu \rho}\, (q \varphi)_\nu  
\nonumber \\[3mm]
&-&
I'_{\nu \rho}\, (q'' \varphi)_\mu - 
I_{\mu \nu}\, (q'' \varphi)_\rho - I_{\mu \rho}\, (q' \varphi)_\nu\},
\label{eq:Pi_1}\\[5mm]
 \Pi^A_{\mu\nu \rho} &=& - \frac{i e^3}{2 \pi^2} 
\{ (q' \varphi q'') \, \tilde \varphi_{\mu \sigma}\pi_{\sigma\nu \rho} + 
 (q'\tilde \varphi I'')_\nu \,\varphi_{ \rho \mu} - 
\frac{1}{2} ((q'' - q') I \tilde \varphi)_\mu\, 
\varphi_{ \nu \rho}  
\nonumber \\[3mm]
&+&
(q'' \tilde \varphi I')_\rho \,\varphi_{ \nu \mu} - 
 (\tilde \varphi I'')_{\rho \nu}\, (q' \varphi)_\mu + 
 (\tilde \varphi I'')_{\mu \nu}\, (q \varphi)_\rho
+ (\tilde \varphi I')_{\mu \rho}\, (q \varphi)_\nu  
\nonumber \\[3mm]
&-&
(\tilde \varphi I')_{\nu \rho}\, (q'' \varphi)_\mu - 
(\tilde \varphi I)_{\mu \nu}\, (q'' \varphi)_\rho - 
(\tilde \varphi I)_{\mu \rho}\, (q' \varphi)_\nu
\}.
\label{eq:Pi_2}
\end{eqnarray}
It is remarkable that the amplitude ${\cal M}$ depends only on two types 
of integrals, $I_{\mu \nu}$ and $\pi_{\mu \nu \rho}$ 
\begin{eqnarray}
I_{\mu \nu} &\equiv& I_{\mu \nu} (q) = - i \pi
\int \frac{d^2 p}{(2 \pi)^2} \, \mathrm{Sp} \{\gamma_\mu
\prl{S}(p - q) \gamma_\nu
\prl{S}(p)\},
\label{I}
\\
\pi_{\mu \nu \rho} &=& - i \pi
\int \frac{d^2 p}{(2 \pi)^2} \, \mathrm{Sp} \{\gamma_\mu
\prl{S}(p - q'') \gamma_\nu
\prl{S}(p) \gamma_\rho \prl{S}(p + q')\},
\label{pi}
\end{eqnarray}
with
$$
\prl{S}(p) = \frac{\prl{(p \gamma)} + m}{\prl{p}^2 - m^2}\, 
\Pi_{-}.
$$
Both types of integrals (\ref{I}) and (\ref{pi}) can be presented in terms 
of analytical functions. 
The integral $I_{\mu \nu}$ can be written as: 
%
$$
I_{\mu \nu} (q) = 
\bigg (
\tilde \Lambda_{\mu \nu} - \frac{q_{\mbox{\tiny $\|$}\mu} \, 
q_{\mbox{\tiny $\|$}\nu}}{\prl{q^{2}}}\bigg )\, \HH,
$$
where
\begin{eqnarray}
&&H(z)=\frac{z}{\sqrt{z - 1}} \arctan \frac{1}{\sqrt{z - 1}} - 1,
\ z > 1,
\nonumber \\
&&H(z) = - \frac{1}{2} \left ( \frac{z}{\sqrt{1-z}}
\ln \frac{1 + \sqrt{1-z}}{1 - \sqrt{1-z}} + 2 -
i \pi \frac{z}{\sqrt{1-z}} \right ), \ z < 1.
\nonumber
\end{eqnarray}
%
The expression for $\pi_{\mu \nu \rho}$ can be presented in the following 
form:
\begin{eqnarray}
&&\pi_{\mu \nu \rho} = \frac{1}
{\prl{q^2} \prl{q'^2} \prl{q''^2}}
\Big [ (q' \tilde \varphi q'')\big\{ 
(\tilde \varphi q)_\mu (\tilde \varphi q'')_\nu 
(\tilde \varphi q')_\rho \prp{\pi}  
\nonumber
\\
&&+(\tilde \varphi q)_\mu (\widetilde \Lambda q'')_\nu
( \widetilde \Lambda q')_\rho H
-(\widetilde \Lambda q)_\mu (\tilde \varphi q'')_\nu
( \widetilde \Lambda q')_\rho H'' 
- (\widetilde \Lambda q)_\mu (\widetilde \Lambda q'')_\nu
( \tilde \varphi q')_\rho H'\big\}
\nonumber
\\
&&+ \prl{(q' q'')} 
( \widetilde \Lambda q)_\mu 
(\tilde \varphi q'')_\nu (\tilde \varphi q')_\rho
(H''-H') 
+
\prl{(q q'')}  
(\tilde \varphi q)_\mu (\tilde \varphi q'')_\nu
( \widetilde \Lambda q')_\rho
(H-H'') 
\nonumber
\\
&&+
\prl{(q q')} 
(\tilde \varphi q)_\mu 
( \widetilde \Lambda q'')_\nu 
(\tilde \varphi q')_\rho
(H'-H)
\Big ], 
\end{eqnarray}
\begin{eqnarray}
\!\!\!\!\!\!\!\!\!\!\!\!\!\!\!\!\!\!\!\!\!\!\!\!\!
\pi_{\mbox{\tiny $\bot$}} &=& H' + H'' + H  
\nonumber
\\[2mm]
\!\!\!\!\!\!\!\!\!\!\!\!\!\!\!\!\!\!\!\!\!\!\!\!\!
&+& 2\frac{\prl{q^{2}} \prl{q'^2}\prl{q''^2} - 
2 m_e^2 [ \prl{q^2} \prl{(q'q'')} H - \prl{q'^2} \prl{(qq'')} H' - \prl{q''^2} \prl{(qq')} H'']}
{\prl{q^{2}} \prl{q'^2}\prl{q''^2} - 4 m_e^2 [\prl{q'^2}\prl{q''^2} - 
\prl{(q'q'')^2}]}.
\end{eqnarray}
where, e.g. $H' \equiv H(4m_e^2/\prl{q'^2})$. The result (\ref{M_fin})  
can be also treated as an effective Lagrangian 
of photon-neutrino interaction in the momentum representation. Moreover it 
can be used to obtain the effective Lagrangians of axion-two photon 
($a \gamma \gamma$) and three photon ($\gamma \gamma \gamma$) interactions by 
making the substitutions 
$$
{\cal M}_{a\gamma\gamma} = 
{\cal M}(g_V=0,\ g_A \frac{G_F}{\sqrt{2} } j_\mu \to \frac{i g_{ae}}{2 m_e} 
\,q_\mu)
$$ 
and 
$$
{\cal M}_{\gamma\gamma\gamma} = 
{\cal M}(g_A=0,\ g_V \frac{G_F}{\sqrt{2} e}  j_\mu \to \ee_\mu)
$$
correspondingly. Here $g_{ae}$ is the axion-electron coupling.

Note also that the amplitude (\ref{M_fin}) 
in the low energy limit, $\omega \ll m_e$, with $g_V = g_A = 1$
coincides with the amplitude obtained in~\cite{Loskutov:1981}.

\section{Contribution of the process $\gamma \gamma \to \nu \bar \nu$
into the neutrino emissivity of the photon gas}

To illustrate a possible application of the result obtained let us 
estimate the contribution of the process $\gamma \gamma \to \nu \bar \nu$
into the neutrino emissivity of the photon gas in strong magnetic field.
It is convenient now to turn from the general amplitude (\ref{M_fin}) to the
partial amplitudes corresponding to definite photon modes with 
polarisation vectors 
$\ee^{(\mbox{\tiny $\|$})}_\alpha = (\varphi q)_\alpha/\sqrt{\prp{q}^2}, \, 
\ee^{(\mbox{\tiny $\bot$})}_\alpha = (\tilde \varphi q)_\alpha/\sqrt{\prl{q}^2}$
(in Adler's notation~\cite{Adler}). 
These amplitudes can be written as
\begin{eqnarray}
\!\!\!\!\!\!\!\!
&&{\cal M}_{\mbox{\tiny $\|$} \mbox{\tiny $\|$}} 
=  i \frac{2 \alpha}{\pi} \, \frac{G_F}{\sqrt{2}} \,
\frac{(q' \varphi q'')(q'\tilde \varphi q'')}
{\prl{q^2} \sqrt{
\prp{q''^2}\prp{q'^2}}}\; [g_V (j \tilde \varphi q) - g_A \prl{(j q)}]\; H, 
\\[2mm]
\!\!\!\!\!\!\!\!
&&{\cal M}_{\mbox{\tiny $\|$} \mbox{\tiny $\bot$}} = - i \frac{2 \alpha}{\pi} \, \frac{G_F}{\sqrt{2}} \,
\frac{1}
{\sqrt{\prp{q''^2}\prl{q'^2}}}\; 
\nonumber
\\
\!\!\!\!\!\!\!\!
&&\times \Bigg \{ g_V \Bigg (
[(j \tilde\varphi q') \prp{(q q'')} + \prp{(j q'')} (q' \tilde \varphi q'') ] H'
-\frac{(j \tilde\varphi q) \prl{(q q')} \prp{(q' q'')}}{\prl{q}^2} H \Bigg )
\nonumber
\\
\!\!\!\!\!\!\!\!
&& -  g_A \bigg (
[ \prl{(j q')} \prp{(q q'')} - \prp{(j q'')} \prl{(q'q'')} ] H'
-\frac{\prl{(j q)} \prl{(q q')} \prp{(q' q'')}}{\prl{q}^2} H \bigg) 
\Bigg \},
\\[2mm]
\!\!\!\!\!\!\!\!
&&{\cal M}_{\mbox{\tiny $\bot$} \mbox{\tiny $\bot$}} = 
- i \frac{2  \alpha}{\pi} \, \frac{G_F}{\sqrt{2}} \,
\frac{1}
{\sqrt{\prl{q'^2}\prl{q''^2}}}\; 
\Bigg \{ \frac{(q' \varphi q'')}{\prl{q^2}}
\Big ( (q' \tilde \varphi q'') 
[g_V (j \tilde \varphi q) - g_A \prl{(j q)}] 
\pi_{\mbox{\tiny $\bot$}}
\nonumber
\\
\!\!\!\!\!\!\!\!
&&+ 
\prl{(q' q'')}
[g_V \prl{(j q)} - g_A (j \tilde \varphi q)] 
(H''-H')
\Big )
 - 
(j \varphi q') H'' [g_V \prl{(q' q'')} - g_A (q' \tilde \varphi q'')]  
\nonumber
\\
\!\!\!\!\!\!\!\!
&&-
(j \varphi q'') H' [g_V \prl{(q' q'')} + g_A (q' \tilde \varphi q'')] 
\Bigg \}
\end{eqnarray}

\begin{figure}[htb]
\centerline{\includegraphics{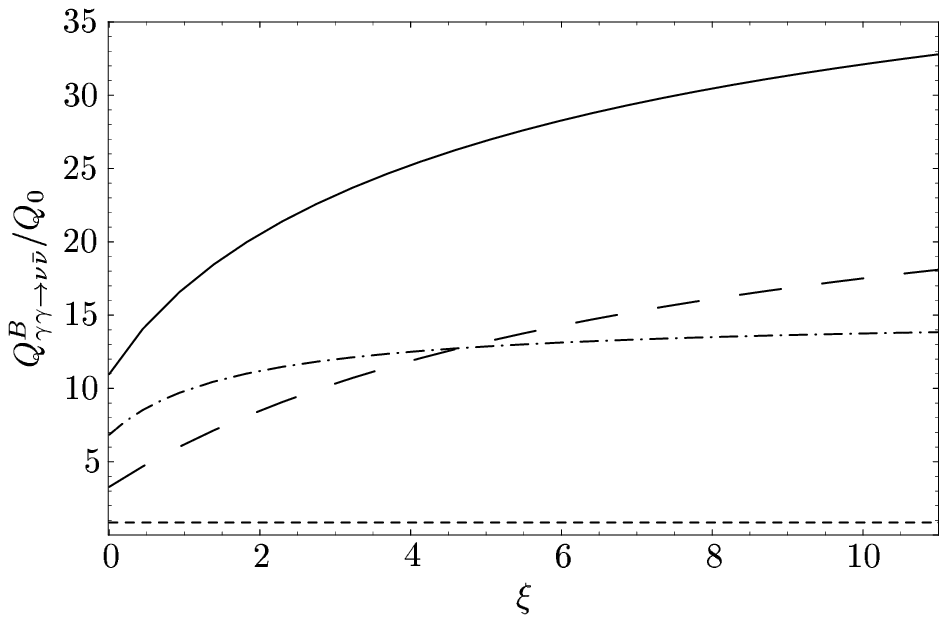}}
\caption{
The low temperature, $T\ll m_e$, neutrino emissivity 
$Q_{\gamma \gamma \to \nu \bar \nu}^{B}$ dependence on the 
parameter $\xi = \frac{\alpha}{3 \pi}\, \frac{B}{B_e}$ for different 
polarisation configurations of the initial photons;
$Q_0 = 10^{8} T_9^{13}\, 
\mbox{\small erg}/(\mbox{\small s}\cdot \mbox{\small cm}^3)$.
Short-dashed, dash-dotted and long-dashed curves correspond to 
$Q_{\mbox{\tiny $\|$}\mbox{\tiny $\|$}}, Q_{\mbox{\tiny $\bot$}\mbox{\tiny $\|$}}, Q_{\mbox{\tiny $\bot$}\mbox{\tiny $\bot$}}$
respectively. 
Solid line depicts the total neutrino emissivity.
}
\label{F2}
\end{figure}

Then the neutrino emissivity (energy carried out by neutrinos from unit 
volume per unit time) can be defined as 
\begin{eqnarray}
Q_{\gamma \gamma \to \nu \bar \nu}^{B} = Q_{\mbox{\tiny $\|$}\mbox{\tiny $\|$}} + Q_{\mbox{\tiny $\bot$}\mbox{\tiny $\|$}} + Q_{\mbox{\tiny $\bot$}\mbox{\tiny $\bot$}},
\end{eqnarray}
\vspace*{-5mm}
\begin{eqnarray}
Q_{\lambda' \lambda''} &=& (2 \pi)^4 g_{\lambda' \lambda''} 
\sum \limits_i \int \vert {\cal M}_{\lambda' \lambda''} \vert^2 
Z_{\lambda'} Z_{\lambda''}
(E_i' + E_i'')
\,\delta^4 (q' + q'' - k' - k'')
\nonumber
\\
&\times& \frac{d^3 q'}{(2\pi)^3 2 \omega'}\, f(\omega')\,
\frac{d^3 q''}{(2 \pi )^3\, 2\, \omega''}\, f(\omega'')\,
\frac{d^3 k'}{(2\pi)^3\, 2\, E_i'}\,
\frac{d^3 k''}{(2\pi)^3\, 2\,	 E_i''}.
\label{Qll}
\end{eqnarray}
Here $E_i', E_i''$ are the energies of the neutrino and antineutrino  
of definite types  $i = \nu_e, \nu_\mu, \nu_\tau$; 
$\omega', \omega''$ are 
the energies of the initial photons; $f(\omega) = [\exp(\omega/T) - 1]^{-1}$
is the photon distribution function at the temperature $T$; the factor 
$g_{\lambda' \lambda''} = 1-\frac{1}{2} \delta_{\lambda' \lambda''}$ is 
inserted to account for the possible identity of the photons in the initial 
state. We would like to note 
that the integration over the phase space of 
initial photons in (\ref{Qll}) has to be performed taking account for 
nontrivial photon dispersion law in the presence of strong magnetic field.
Moreover, it is necessary to take into consideration the 
large radiative corrections in strong magnetic field which are reduced to the 
wave-function renormalization factors $Z_{\lambda'}$ and $Z_{\lambda''}$  in 
Eq.~(\ref{Qll}). In the case of low temperature, $T \ll m_e$ the  
dispersion law of the photon with polarization vector 
$\ee^{(\mbox{\tiny $\bot$})}_\alpha$ and corresponding 
wave-function renormalization factor 
$Z_{\mbox{\tiny $\bot$}}$ could be presented in the following form
\begin{eqnarray}
\omega^2 = \frac{\prp{q}^2}{(1 + \xi)} + q_3^2, \qquad
Z_{\mbox{\tiny $\bot$}} \simeq \frac{1}{1 + \xi},
\label{deqLE}
\end{eqnarray}
where $\xi = \frac{\alpha}{3 \pi} \, \frac{B}{B_e}$ is factor 
characterising magnetic field influence. The photon with polarization 
vector $\ee^{(\mbox{\tiny $\|$})}_\alpha$ has 
almost vacuum dispersion law and wave-function renormalization factor 
in this limit, $q^2 \simeq 0, \,\, Z_{\mbox{\tiny $\|$}} \simeq 1$.
It is useful also to present here the element of the momentum space 
\begin{eqnarray}
d^3 q = (1 + \xi)\, \omega^2\, d \omega\, d \varphi\, d y,
\qquad  y = \cos \theta \sqrt{1 + \xi} / \sqrt{1 + \xi \cos^2 \theta},
\end{eqnarray}   
where $\theta, \phi$ are the polar and the azimutal angles.
All of these facts lead to the rather complicated dependence of 
$Q_{\gamma \gamma \to \nu \bar \nu}^{B}$ on 
magnetic field strength despite the amplitude (\ref{M_fin}) does not contain this 
dependence. In the case of the arbitrary parameter $\xi$ the value of the 
contribution into neutrino emissivity of 
the photon gas can be found only numerically, but when 
the strength of  magnetic field  is not too  large, $eB \ll m_e^2 / 
\alpha$, one can obtain the analitical expressions for $Q_{\lambda' 
\lambda''}$:
\begin{eqnarray}
Q_{\mbox{\tiny $\|$}\mbox{\tiny $\|$}} &=& \frac{8\,\alpha^2 G_F^2}{8505 \,\pi^3 \,m_e^4}\,\, T^{13}\,  
\bigg[  {\bar g}_V^2 \pi^2 \left(7 \pi^2 \zeta(5) + 330\, \zeta(7)\right)
\nonumber
\\
&+& {\bar g}_A^2 \left( 9 \pi^4 \zeta(5) + 
 70 \pi^2 \zeta(7) + 420\, \zeta(9) \right )\bigg]  
\nonumber
\\
&\simeq& 8.5 \cdot 10^7 \, T_9^{13}\,
\frac{\mbox{erg}}{\mbox{s}\cdot \mbox{cm}^3 }\, ,
\label{Q11}
\\[35mm]
Q_{\mbox{\tiny $\|$}\mbox{\tiny $\perp$}} &=& \frac{32\,\alpha^2 G_F^2}{127575\, \pi^3 \, m_e^4}\, T^{13}\,
\bigg[  {\bar g}_V^2 24 \pi^2 \left(14 \pi^2 \zeta(5) + 285\, 
\zeta(7)\right) 
\nonumber
\\
&+&   {\bar g}_A^2 \left( (309 \pi^2 + 20) \pi^2 \zeta(5) 
+ (1410 \pi^2 + 315) \zeta(7) + 38430\, \zeta(9) \right )\bigg] 
\nonumber
\\
&\simeq& 6.9 \cdot 10^8 \, T_9^{13}\,
\frac{\mbox{erg}}{\mbox{s}\cdot \mbox{cm}^3 }\, ,
\label{Q12}
\\[2mm]
Q_{\mbox{\tiny $\perp$}\mbox{\tiny $\perp$}} &=& \frac{8\,\alpha^2 G_F^2}{127575\, \pi^3 \, m_e^4}\, T^{13}\,
\bigg[  {\bar g}_V^2 \pi^2 \left((609 \pi^2 - 50) \zeta(5) + 10110\, 
\zeta(7)\right) 
\nonumber
\\
&+&   {\bar g}_A^2 \left( (879\, \pi^2 + 50) \pi^2 \zeta(5) 
+ 5490 \pi^2 \zeta(7) + 56700\, \zeta(9) \right )\bigg] 
\nonumber
\\
&\simeq& 3.3 \cdot 10^8 \, T_9^{13}\,
\frac{\mbox{erg}}{\mbox{s}\cdot \mbox{cm}^3 }\, .
\label{Q22}
\end{eqnarray}
Here $T_9$  is the temperature in units of $10^9 K \simeq 0.17 m_e$,
the effective constants~${\bar g}_V^2 \simeq 0.93, \, {\bar g}_A^2 \simeq 0.75$ 
are summarized over all channels of the neutrino production, 
$\nu_e, \nu_\mu,\nu_\tau$.
The total neutrino emissivity in this case is 
\begin{eqnarray}
Q_{\gamma \gamma \to \nu \bar \nu}^{LT}\simeq
1.1 \cdot 10^9 \, T_9^{13}\,
\frac{\mbox{erg}}{\mbox{s}\cdot \mbox{cm}^3 }\,.
\label{Qne}
\end{eqnarray} 
We also have made the numerical calculation of 
the neutrino emissivity caused by the process $\gamma \gamma \to \nu \bar \nu$.
In the low temperature limit $T \ll m_e$ our result is represented in 
Fig.~\ref{F2} where the emissivity $Q_{\gamma \gamma \to \nu \bar \nu}^{B}$ 
is depicted as a function of the 
parameter $\xi$. 
%

The result presented by formular (\ref{Qne})  and in Fig.~\ref{F2} should be compared with the contributions into 
the neutrino emissivity
of the process $\gamma \gamma \to \nu \bar \nu$ caused by the another 
mechanisms. For instance, the emissivity due to the finite neutrino mass 
is~\cite{Dodelson:1991}
\begin{eqnarray}
Q_{\gamma \gamma \to \nu \bar \nu}^{m_\nu}\simeq
1.4 \cdot 10^{-4} \, T_9^{11}\,
\frac{\mbox{erg}}{\mbox{s} \cdot \mbox{cm}^3 }\, \left (\frac{m_\nu}{1\, \mbox{eV}}\right )^2.
\label{Qmn}
\end{eqnarray}  
On the other hand, in the case of non-locality of the weak interaction, 
investigated in Ref.~\cite{Dicus:1993}, one can estimate the emissivity, which is 
suppressed by the factor $(m_e/m_W)^4$:
\begin{eqnarray}
Q_{\gamma \gamma \to \nu \bar \nu}^{NL}\simeq
9.9 \cdot 10^{-10} \, T_9^{13}\,
\frac{\mbox{erg}}{\mbox{s} \cdot \mbox{cm}^3 }\,.
\label{QNL}
\end{eqnarray}
It's obvious that the field-induced mechanism of the reaction 
$\gamma \gamma \to \nu \bar \nu$ strongly dominates all the other 
mechanisms.

It is interesting also to compare our result  with the previous 
calculations of the neutrino emissivity due the process 
$\gamma \gamma \to \nu \bar \nu$ in the weak and strong magnetic fields.
Taking into account the remark by authors 
of~\cite{Dicus:2000} one could obtain from~\cite{Shaisultanov:1998} the following estimation for the 
neutrino emissivity $Q_{\gamma \gamma \to \nu \bar \nu}^{B}$ in the weak 
magnetic field limit, $B \ll B_e$:
\begin{eqnarray}
Q_{\gamma \gamma \to \nu \bar \nu}^{B} \simeq   0.3 \cdot 10^9\, T_9^{13} \left(\frac{B}{B_e}\right)^2
\frac{\mbox{erg}}{\mbox{s} \cdot \mbox{cm}^3 }.
\label{QneSh}
\end{eqnarray}
It is evident that neutrino emissivity is substantially enhanced in 
strong magnetic field in comparison with its counterpart in the 
weak magnetic field case.

In the limit $B \gg B_e$ the contribution of the process 
$\gamma \gamma \to \nu \bar \nu$  into the neutrino emissivity 
was previously  studied in~\cite{Loskutov:1981}, from which the following 
estimation could be  obtained:
\begin{eqnarray}
Q_{\gamma \gamma\to \nu \bar\nu}^{B} \simeq 0.7 \cdot 10^8\, T_9^{13}
\frac{\mbox{erg}}{\mbox{s} \cdot \mbox{cm}^3 }.
\label{QneSk}
\end{eqnarray}
This result doesn't depend on the magnetic field strength $B$ and it
is at least ten times less than our result presented in Fig.~\ref{F2}. 
In our opinion, the  authors~\cite{Loskutov:1981} wrongfully didn't take into 
account photon dispersion and wave-function renormalisation in strong 
magnetic field.

Let us note that in the presence of the magnetic field one more contribution 
into neutrino emissivity due to the process $\gamma \to \gamma \nu \bar 
\nu$ is possible. We would like to emphasize that it is the nontrivial  
dispersion law of a photon in the magnetic field that makes this reaction to be 
kinematically allowed. To our knowledge the process 
$\gamma \to \gamma \nu \bar \nu$ has not been studied so far. Therefore 
it is interesting to compare the contributions into the neutrino emissivity 
from the $\gamma  \gamma \to \nu \bar \nu$ and $\gamma  \to \gamma  \nu \bar 
\nu$ channels. 

\begin{figure}[htb]
\centerline{\includegraphics{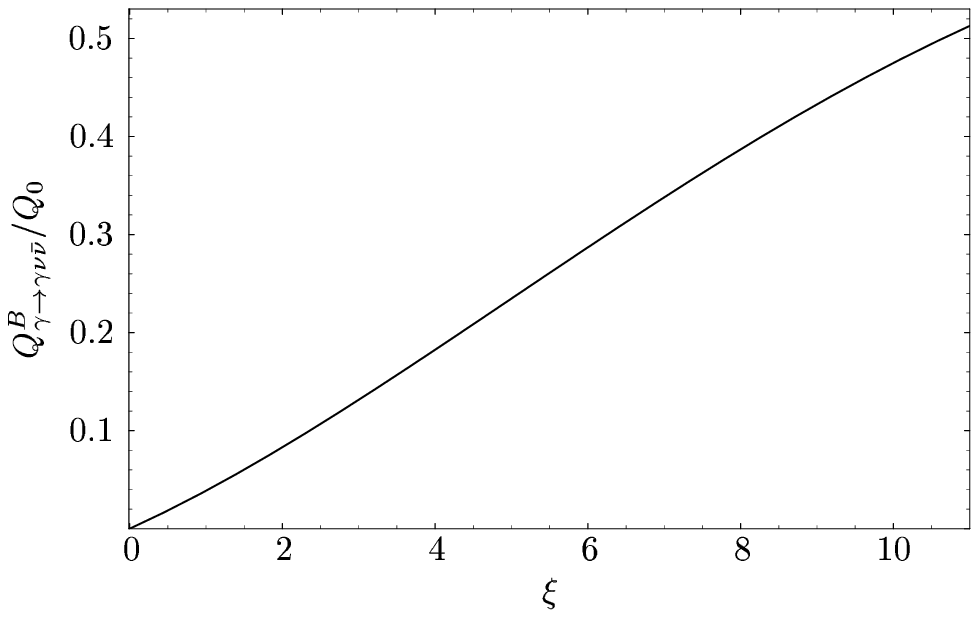}}
\caption{
The low temperature, $T\ll m_e$, neutrino emissivity 
$Q_{\gamma \to \gamma  \nu \bar \nu}^{B}$
dependence on the parameter $\xi$.
}
\label{F3}
\end{figure}

To obtain the emissivity by the process 
$\gamma  \to \gamma  \nu \bar \nu$ one needs to make the replacements
$f(\omega) \to (1 + f(\omega))$  and $q \to - q $ for one of the photons 
in (\ref{Qll}).
The  analysis of the process kinematics shows that only one transition,
$\prl{\gamma} \to \prp{\gamma} \nu \bar \nu$, gives the contribution into 
neutrino emissivity. The dependence of the neutrino emissivity due to the 
process $\gamma \to \gamma \nu \bar \nu$ on the magnetic field strength is 
depicted in Fig.\ref{F3}.
As is seen from both figures the contribution of the 
process $\gamma \to \gamma \nu \bar \nu$ into the neutrino emissivity 
turns out to 
be small in comparison with analogous contribution due to the reaction  
 $\gamma  \gamma \to\nu \bar \nu$ in the case of not too strong magnetic 
field, $B\ll 10^5 B_e$.

In the high temperature limit, $T \gg m_e$, the main contribution into 
neutrino emissivity arises from the photons with parallel momenta squared are 
in the region $m_e^2 \ll \prl{q}^2 \ll eB$. In this case one can use 
vacuum dispersion law for both photon modes. We obtain the following 
estimation for neutrino emissivity in this limit:
\begin{eqnarray}
Q_{\gamma \gamma \to \nu \bar \nu}^{HT}  \simeq  2.7 \cdot 10^{18}  \left( \frac{T}{m_e} 
\right)^9
\frac{\mbox{erg}}{\mbox{s} \cdot \mbox{cm}^3 }\,. 
\label{QnHT}
\end{eqnarray}
Let us compare this result with contribution into neutrino emissivity from 
process $\gamma \to \nu \bar \nu$. In strong magnetic field this 
reaction was studied in~\cite{Kuznetsov:1998}. It was shown that 
the process  $\gamma \to \nu \bar \nu$ is kinematically allowed in the 
region $\prl{q^2} > 4 m^2$. This lad to the suppression of the process in 
the low temperature limit by factor $\prl{q}^2 \simeq 4m_e^2$. In high 
temperature limit authors note that the main contribution into neutrino 
emissivity is defined by the vicinity of the point $\prl{q^2} = 4 m_e^2$. 
However, the detailed analysis shows that the main contribution arises 
from the region $m_e^2 \ll \prl{q}^2 \ll \beta$. In this case the
estimation of the  neutrino emissivity coincides with the formula 
obtained by authors in the limit  $T^2 \gg eB, m_e^2$:
\begin{eqnarray}
Q_{\gamma \to \nu \bar \nu}^{HT} \simeq
0.40 \cdot 10^{18} \left(\frac{T}{m_e}\right)^5
\left(\frac{B}{B_e}\right)^2 
\frac{\mbox{erg}}{\mbox{s} \cdot \mbox{cm}^3 }\,. 
\label{Qgnn}
\end{eqnarray}
Then the ration of the emissivities (\ref{QnHT}) and (\ref{Qgnn}) is 
\begin{eqnarray}
R = \frac{Q_{\gamma  \to \nu \bar \nu}^{HT}}
{Q_{\gamma \gamma\to \nu \bar \nu}^{HT}} \simeq  0.15 \left(\frac{B}{T^2}\right)^2.
\label{R_Q}
\end{eqnarray}
It is seen that the contribution into neutrino emissivity of the process 
$\gamma \to \nu \bar \nu$ is always larger then 
$\gamma \gamma \to \nu \bar \nu$ reaction contribution, because $B \gg T^2$ in 
strong magnetic field limit. 
For example, $R \simeq 5.86$  for $T = 2$ MeV and $B = 100 B_e$. 
Nevertheless, we can see that these quantities have the comparable values.

In summary, we have investigated the two-photon two-neutrino processes 
in the presence of strong magnetic field. The amplitude of the reaction 
$\gamma \gamma \to \nu \bar \nu$ is 
obtained in a general case when the photons are not assumed to be on the 
mass shell. Therefore it can be treated as an effective Lagrangian of 
photon-neutrino interaction. We have also  calculated the contribution 
into the neutrino emissivity due to the reactions 
$\gamma \gamma \to \nu \bar \nu$ and $\gamma \to\gamma  \nu \bar \nu$ 
taking into account  the photon dispersion and wave function 
renormalisation in strong magnetic field. 
We stress that in spite of the independence of the amplitude  on the 
magnetic field strength $B$, the
emissivity essentially depends on $B$. The comparison of our results with 
the other inducing mechanisms of the reaction $\gamma \gamma \to \nu \bar \nu$  
shows that strong magnetic field catalyses this process. 
Moreover it seems that 
the processes under consideration are the most dominating photon-neutrino 
reactions.

\vspace{7mm}

\noindent 
{\bf Acknowledgements}  

\noindent
M.C. expresses his deep gratitude to the organizers of the  
XXXI ITEP Winter School of Physics for warm hospitality.

This work was supported in part by the Russian Foundation for Basic 
Research under the Grant N~01-02-17334, 
by the Ministry of Education of Russian Federation under the 
Grant No.~E02-11.0-48 and by 
Ministry of Industry, Science and Technologies of Russian Federation 
under the Grant No.~NSh-1916.2003.2.


\newpage

\end{document}